# Improving Brain Magnetic Resonance Image (MRI) Segmentation via a Novel Algorithm based on Genetic and Regional Growth

Javadpour A.[1], Mohammadi A.[1]*

[1]Neuroscience Research Center, Baqiyatallah University of Medical Sciences, Tehran, Iran

## ABSTRACT

**Background:** Regarding the importance of right diagnosis in medical applications, various methods have been exploited for processing medical images solar. The method of segmentation is used to analyze anal to miscall structures in medical imaging.

**Objective:** This study describes a new method for brain Magnetic Resonance Image (MRI) segmentation via a novel algorithm based on genetic and regional growth.

**Methods:** Among medical imaging methods, brains MRI segmentation is important due to high contrast of non-intrusive soft tissue and high spatial resolution. Size variations of brain tissues are often accompanied by various diseases such as Alzheimer's disease. As our knowledge about the relation between various brain diseases and deviation of brain anatomy increases, MRI segmentation is exploited as the first step in early diagnosis. In this paper, regional growth method and auto-mate selection of initial points by genetic algorithm is used to introduce a new method for MRI segmentation. Primary pixels and similarity criterion are automatically by genetic algorithms to maximize the accuracy and validity in image segmentation.

**Results:** By using genetic algorithms and defining the fixed function of image segmentation, the initial points for the algorithm were found. The proposed algorithms are applied to the images and results are manually selected by regional growth in which the initial points were compared. The results showed that the proposed algorithm could reduce segmentation error effectively.

**Conclusion:** The study concluded that the proposed algorithm could reduce segmentation error effectively and help us to diagnose brain diseases.

## Keywords

Brain Magnetic Resonance Image, Segmentation, Regional Growth, Genetic Algorithm

## Introduction

Medical imaging and Electroencephalography (EEG) have important applications in different medical fields and is used to primary evaluation of neurological disorders. Among advances in science and technology due to high efficiency of calculations, medical imaging is more highlighted, because of its ability to illustrate human's internal organs. Medical imaging is leading in health treatment, medical investigations and education. This medical realm provides new opportunities to improve health care through tech no logy and to monitor internal organs non intrusively and to investigate and to interpreter which organs are healthy and which are not [1, 2]. Alzheimer is a pro-gressive neurodegenerative disease which is the main cause of concern among old people [3, 4]. This disease is identified by brain atrophy and microscopically irritated by light. Based on ADI report in 2009, about 36 million people are suffering Alzheimer in the world and the number is doubled every 20 years i.e. this number will reach 66 to 115 million people in 2030 and 2050, respectively. The rate of this rising process is higher in developing countries. Annual expense for treatment of such patients was estimated as much as 604 billion dollars in 2010 [5]. Regarding the growing number of old people in some Asian developing countries, it is estimated the number of people older than 60 will reach 700 million in 2025 resulting in considerable reduction in number of labors . On the other hand, this event will accompany old-age-related diseases such as Alzheimer. There is no comprehensive information about the number of people suffering Dementia and Alzheimer but based on Iranian Alzheimer Association, this number is about 800 thousand. Regarding the growing number of old people in the future, growing suffering people are much expected. This would accompany higher cost of keeping and management of such people as well as higher energy consumption for this purpose. During recent decades, the evaluation of neuropsychology is considered as one of the key approaches to identify Alzheimer, which is used to register and rate the reduction in various realms of cognitive functions [6, 7].

*Corresponding author: A. Mohammadi, Ph.D., Assistant Professor, Neuroscience Research Center, Baqiyatallah University of Medical Sciences, Tehran, Iran Email: ar.mohammadi@bmsu.ac.ir

In this paper, a comprehensive study on types of cognitive differences pertaining to different regions is performed in a group of patients suffering Alzheimer based on common neurologic examinations on Alzheimer patients. Physical differences in speech of these patients have been identified. Disorders have considerable impacts on their daily life and activities. So, the evaluation of psychology to identify and monitor different regions of prefrontal cortex of Alzheimer in suffering patients is crucial. Another advantage pertains to specify the harmful effects of pharmaceutical treatments in this field along with the evaluation of psychology nerve.

## Material and Methods

### Medical Image Segmentation

As our knowledge about their lotion between various brain disorders and anatomical brain deviation grows, brain segmentation method is exploited as the first step in diagnosis and analysis. For example, Alzheimer and sclerosis are disorders that can be studied based on deviation in brain structure in this paper; we study methods that are used in brain MRI segmentation imaging and suggest improvements. Medical image segmentation is crucial in most medical applications such as planning surgery, after surgery evaluation, discovering anomalies and so on [8].

Complicated strictures are often included in medical images and precise segmentation of these images is necessary for appropriate diagnosis. Precise segmentation of brain images is crucial to identify tumors, timidity and dead or corrupted tissues as well as appropriate and precise discovery of such tissues in diagnostic methods. Magnetic resonance imaging (MRI) is also an important imaging to discover and identify animal variations in various parts of brain stages. MRI is a well-known method to record brain images with high contrast. MRI parameters can be tuned so that gray material surfaces pertaining to various tissues and wide variety of neurological anomalies are discovered. MRI images have better contrast then computer topography (CT) method. Hence, this method is exploited in most investigations on medical imaging investigations [9]. Flying brand structure by magnetic resonance imaging (MRI) is central in neurology and has various applications such as performing applied activation scheme on brain anatomy, studying brain growth and neural anatomical variability analysis in normal brain [10]. Segmentation of brain image is also useful in the diagnosis of anomaly ties pertaining to deviation and corruption of neuron and psychology, evaluation of treatment process and planning for surgery (Figure 1). There are many methods for automatic and semi- and auto- segmentation of images but most of them are not successful due to unknown noises, poor image contrast and poor boundaries [11].

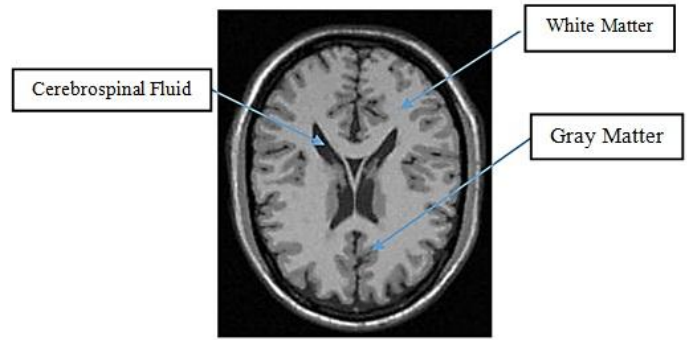

**Figure 1:** Digital Image of Brain

Application of image processing techniques has had rising growth during recent years. Nowadays, recording and storing medical images are performed digitally; although, interpretation and details of images are still time consuming. This issue becomes important, in particular, for some regions of the images where color and shape are animal and should be diagnosed by neurologist for more details and investigation [12]. Image segmentation is part of some warning parts in many image processes and computer image applications. The aim of image segmentation is derision of an image into components based on desired criterion for further processing. Regarding the importance of precise diagnosis in medical applications, various methods have been exploited for image processing, so far MRI images have high contrast and are used for diagnosis. Segmentation of resulting images is very important. Due to the growing importance of medical images for critical medical decisions, exploring technology to cancel human errors and getting the best image explains the necessity of studies in such a realm. In this paper, we investigate current techniques for MRI image segmentation and follow regional growth method for MRI image segmentation. This method can also be used as an input for image analysis tools and software.

## Comparing Results of Implemented Methods

As the size and number of medical images increase, using computers for faster processing and analysis becomes inevitable .In particular, computer algorithms are considered as a vocal part to describe anatomical structures in radiology automatization. Such algorithms are called image segmentation algorithms playing an important role in biomedical imaging applications such as determination of tissue volume [13], diagnosis [13], determination of pathological locations [14], studying anatomy [15] , planning treatment [15], correcting partial volumetric data pertaining to and comprehensive surgery using computers [13]. For the reset, we explain some common methods offered by recent articles about medical image segmentation. So far, several methods have been developed. Some of them are 1-there is hold determination 2-region growth 3-classifiers 4-clustering 5-FCM 6 and active control. FCM was first introduced by [16] and improved by [17]. FCM algorithm is poor in working on important properties of images because neighboring pixels are interactive causing strong perturbation sensitivity. To overcome such a weakness, a new classification algorithm called PCM is introduced in [18].

PCM moderates total pressure of columns in a fuzzy matrix and in traduces a probable partition matrix. PCM pays a cost on ignoring disorder points. First, it is sensitive to ignoring disorder points. Second, it is sensitive to disorder and often leads to the problem of adaptive classification. On the other hand, potential membership is sensitive to selection of additional PCM parameters [19]. Lu and Cowarkars [20] introduced an improved method for FCM to solve the problem of MRI classification. In this algorithm, statistical histogram of gray surface of image in KFCM algorithm is included for algorithm speed up as well as special information of image by KNN algorithm based on core methods in this method, spatial membership matrix between pixels and cones are constructed and then it is transcribed by membership matrix in traditional FCM algorithm which leads to some limitations. In what follows, we consider some methods for MRI segmentation [21].

### Threshold Determination Method

Volume and intensity distribution are usually very complicated in medical images and threshold methods often fail. In most cases, threshold determination method is combined with other methods. One of these methods is segmentation based on threshold. In threshold method, a region is separated based on pixels with similar intensities. This method provides boundaries that separate objects from background based on their contrasts. Threshold method provides a binary output image from a gray image [22].

One of these methods is called general thresholding in which it is assumed that pixels of target and background have differences in gray surface and threshold is selected so that the target is separated from background. Another thresholding method is adaptive thresholding. This method is used when thresholding is not constant and the threshold is specified based on the location of target. To select the threshold, there are various methods including selection of threshold with the help of histogram, clustering and iteration. In threshold determination methods, measurable images are divided into two types of intensities, it tries to separate desired classes by the value of intensity called threshold. Segmentation is resulted by classification of all pixels with higher values of threshold in one group and the rest of pixels in another group [23]. Threshold determination is a simple but effective tool to segment images in which various topologies have various separating intensities or other measurable properties. One segment is farmed interactively; however, there ore automatic methods. Threshold determination is often considered as early stage sequential image process. One of basic red tritons of this method is that in the simplest case of thresholding, only two classes are created that could not be used for multi-channel images. Furthermore, determination of threshold usually does not consider spatial properties of an image. This is why threshold determination method is sensitive to non-uniformity of noise intensity which is often present in MRI images [23]. For these reasons, diversity in threshold determination to segment medical images is proposed which records based on local intensity and data stream [24] threshold techniques are presented in [22].

### Region Growth Method

In this method, region growth operation should be performed in each segment and there is the same issue of appropriate threshold determination for smoothness as threshold determination method. This method begins with the selection of one or more granular points and depending on the number of granular points, the same regions are specified. The similarities between pixels are formulated. This is a criterion based on the gray scale of pixels. Two pixels are similar in which absolute difference of their gray scale is lower than threshold value. Then, region growth operation or monitoring and control of points enclosed in side granular points begin. If a point is similar enough to granular points, it will belong to that granular point [23]. For each region, the procedure continues until all available points in an image are covered. Region growth method is a technique to extract a region of an image that is connected based on some predefined criteria. These criteria could be based on information about intensity and/or image edges. In the simplest form, this method needs one grain that is selected by operator manually and all pixels connected to the grain are extracted with the same intensity initially [25]. This is shown in Figure 2 (B) where region growth method is used to extract one of its structures. As in threshold determination method, this method is not often used lonely. Rather, it is used in a series of image processing operations, in particular, to describe small and simple structures such as tumors and injuries [5]. Hence, for each region to be extracted, one grain needs to be implanted. Separation algorithms depend on region growth method but do not need granular implant. Region growth method is sensitive to noise because it results in holes of disconnection. In contrast, negligible volumetric effects can lead to separated regions connected to each other. To overcome such problems, homo-topic growth algorithm is suggested to keep topology of initial and extracted regions.

### Artificial Neural Networks

Artificial neural networks (ANN) are processing nodes that simulate biological information. Each node in ANN is capable of executing primary calculations. Learned information is gained by adaption of assigned weights for connections among nodes. ANNs are an example of machine learning and could use in various types of image segmentation. ANN can be used in inside monitoring methods such as classification method and also in change formable models. Due to huge number of interconnections used in neural network, spatial information could be used readily along with classification methods. Although ANN is inherently symmetric, processing is usually in series and hence its potential capabilities are decreased [25].

### Clustering Method: Cluster is a similar collection of data

In clustering method, data are divided into clusters in which the similarities between data inside each cluster are maximized and between different clusters are minimized. Clustering algorithms essentially play their roles as classification methods. Hence, these methods are called unsupervised methods. In order to compensate for the lack of educational data, classification methods are repeated between image segmentation and determination of properties of each class. On the one hand, classification methods self-educate accessible data and put these data accessible to the user on the other hand. Three common classification algorithms are k-means or ISO DATA [9] and Expectation Maximization [EM] algorithms [21].

K-means clustering algorithm classifies data by calculating iterative average of intensity for each class and image segmentation through classification of each pixel of a class or the closet average [26]. Figure 3 shows the result of executing k-means algorithm in some regions of brain MRI, Figure 3 (A). The numbers of classes are assumed 3 including; cerebrospinal fluid, gray matter and white matter (from dark gray to white, respectively). FCM algorithm is a generalization of k-means algorithm and provides the possibility of rough classifications based on fuzzy collection theories. EM algorithm applies the same values as clustering algorithms with the following important assumption that data follow the Gaussian combination model [27]. One of clustering methods used in MRI image segmentation is Fuzzy Clustering Method (FCM). This clustering method as introduced in [5] is based on minimization of target function. FCM optimizes the target function through updating membership function and center of clusters. This optimization process continues until reaching the threshold. FCM is widely used in medical image segmentation. Convergence time and sensitivity to noise issues with this algorithm and some development have been made to solve these issues [28].

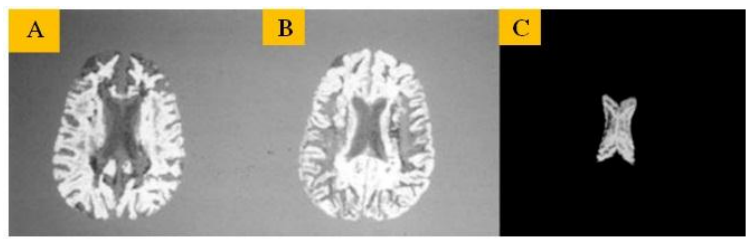

**Figure 2:** MRI Segmentation into Gray Matter, White Matter and Cerebrospinal Fluid Components Using Region Growth Method

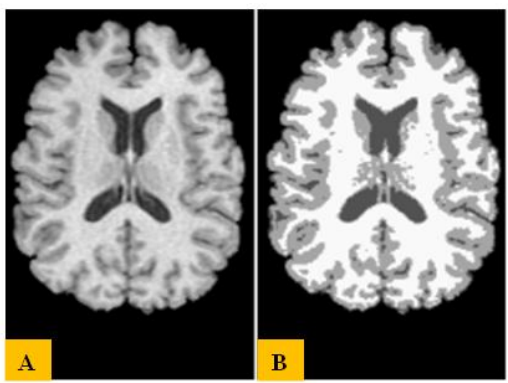

**Figure 3:** Applying k-means Algorithms on Brain Image (A): Original Image, (B): Resulted Image

### Watershed Method

Watershed segmentation technique is based on slope. Different values of slopes are considered as various heights. By creating a hole in each minimum and immersing in water, water level will be local maxima. When two sections of water reach each other, a barrier will be created and water will rise to a level where all points on a map immerse in. Image is segmented by barriers. These barriers are called watershed, and segmented regions are called boundary catchment area. Fast executing method is proposed [29]. There is the problem of segmentation.

### Graph-cut Method

This method is another image segmentation method that operates based on minimizing an energy function. In this method, pixels of an image are labeled; image is segmented into desired segments of application. An image is converted to a graph where each pixel of image is considered as a node of the graph. Then, some criteria are applied to check similarities such as image color or neighborhood of pixels [11].

### Region Growth Method

Region growth method is based on similarity or homogeneity of neighboring pixels. In this method, an image is segmented into separate regions based on a criterion to check homogeneity. Advantages and disadvantages of this method are listed in Table 1 [5]. Pixels inside each region have similarity based on a particular criterion such as color, intensity etc. Region growth method is one of the methods relying on simple region that checks whether selected pixels around primary grain belong to the region or not. Methods relying on histogram for image segmentation only consider gray surface distribution of an image; whereas, region growth methods consider this point whether neighboring pixels comparable gray surface. Region growth-based methods operate as follow:

1. Some primary grains are considered to start the algorithm.

2. Region growth starts from these grain and pixels similar to primary pixels which are added to the region.

**Table 1:** Advantages and Disadvantages of Region Growth

| Advantages |
|---|
| Regions having desired specifications are correctly separated. |
| Good segmentation results for image have distinctive edge. |
| It is simple and develops only by some initial grains. |
| One can select initial grains and desired criterion. |
| Several criteria can be used simultaneously. |
| There is good performance with respect to noise. |
| **Disadvantages** |
| It is time-consuming. |
| This method may lead to additional segmentation or whole creation of image having high fluctuations in intensity. |

3- When region growth stopped, next grain is considered and next region growth starts from there.

4. These steps continue until all image pixels belong to region.

Since proposed method in this article is based on [22], the details are explained in 3 steps (Figure 4):

### Step 1

Selecting primary grains for each region to start algorithm, initial grains need to be given manually. In manual method, algorithm starts after choosing initial grains by user (N and Nair n.d.). Same developments have been made to select initial grains automatically. One of these developed algorithms is random step algorithm to find initial points. Another method to select points is using image histogram. In this method, peaks in histogram are used [30]. First, histogram of an image is extracted. Then, this histogram is segmented regions and final step is selecting points (grain). Pixel X is considered as initial pain to start, if its gray scale lies in a meaningful interval representing the band which is in it. If more than one pixel in a meaningful interval represents a band, then both pixels belong to a class [31]. Another method is based on algorithm relying on genetic algorithm and fuzzy classification. This algorithm operating as fallows; first clustering is made exploiting fuzzy clustering method. Cluster centers (c) and degree of membership (m) is specified by this algorithm. These appropriate values for these parameters are given genetic algorithm so that target function is minimized [24].

### Step 2

Determining a criterion for similarity of regions after specifying initial paints, it is time to select a criterion for similarity of regions. This criterion is used to check similarity between a new pixel and pixels of region that specifies dependency of new pixel to a region.

### Step 3

As mentioned before, region growth to specify initial grains, this method selects initial grains so that it leads to the least errors. After specifying initial grains to start algorithm and similarity criterion that is considered for pixels and regions, region growth begins. As mentioned before, region growth selects initial grains so that it leads to the least errors. Using this method, one can reach an appropriate position for initial grains to start the algorithm. After specifying initial grains, a criterion is exploited for region growth. In this study, standard deviation is used as a criterion because it has good capabilities to identify pixels belonging to a region. 8 neighbors for each pixel are considered and searching begins from initial grain and neighboring pixels are checked and if they belong to that class then they are added. By searching initial grains, their neighboring is selected according to figures 5, 6 and 7 [22].

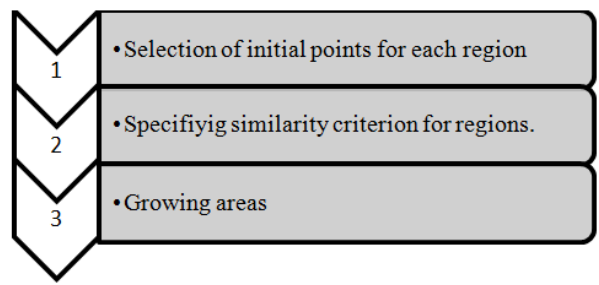

**Figure 4:** Steps of Growing Regional Algorithm

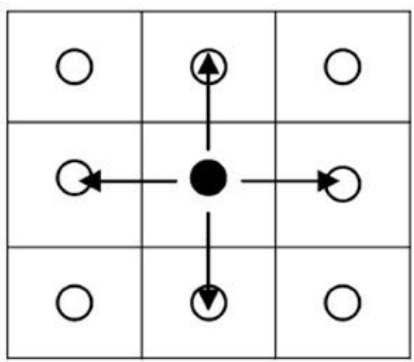

**Figure 5:** Four-neighbor Growth

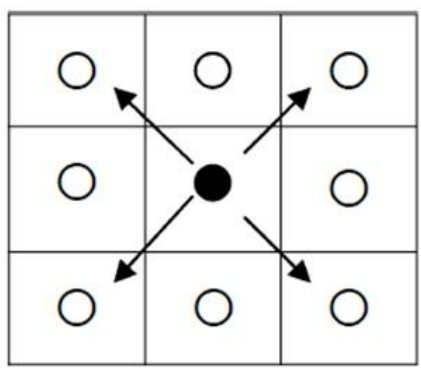

**Figure 6:** Diagonal Four-neighbor Growth

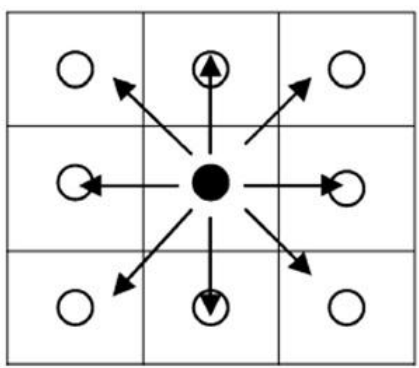

**Figure 7:** Eight-neighbor Growth

## Proposed Method

As mentioned before, region growth method is used for brain image segmentation. This method is rather simple and has the possibility to select initial grains and similarity criterion to reach the desired segmentation. We used automatic method and initial grains are selected automatically. To select initial grains pertaining to gray matter, white matter and cerebrospinal fluid regions of brain in region growth method, we used an automatic method relying on genetic algorithm. This searching continues until first class is complete and then similar procedure is repeated for next classes. Data used in this paper are taken from Brain Web database [32]. MRI images along with their tissue images (gray matter, white matter and cerebrospinal fluid) are available separately.

Figure 8 shows one sample image from this database. Gray matter, white matter and cerebrospinal fluid images pertaining to Figure 8 are available individually in this Web dataset (Figure 9).

### Proposed Simulation

Initial pointes selection to start region growth segmentation was performed using genetic algorithm. So, first we explain this algorithm and then we present some details about used region growth method. After that, we explain the procedure. Genetic algorithm is an inspiration of genetics and Darwin's theory of evaluation, based on tap survival or natural selection. One common application of genetic algorithm is using it as optimization function. Genetic algorithm is a useful tool in pattern recognition, property selection, image identification and machine learning. In genetic algorithms, procedure for genetic evaluation organisms is simulated. In a genetic algorithm, populations of people survive based on their preference. People with superior capabilities undergo changes for marriage and reproduction. Then, children with higher capabilities are reproduced after several generations. In genetic algorithm, each individual of population is introduced as a chromosome. Chromosomes evolve during several generations. In each generation, chromosomes are evaluated and have the possibility to survive and reproduce based on their quality. Reproduction in genetic algorithm is performed through mutation and crossover operators. Superior parents are selected based on a fitness function. In each step of genetic algorithm, execution one class of parents of searching space is processed randomly: a sequence of characters is assigned to each point. Resulting sequences are decoded to give new points in searching space. Finally, their probability to contribute in next step is determined based on the value of target function in each point [24].

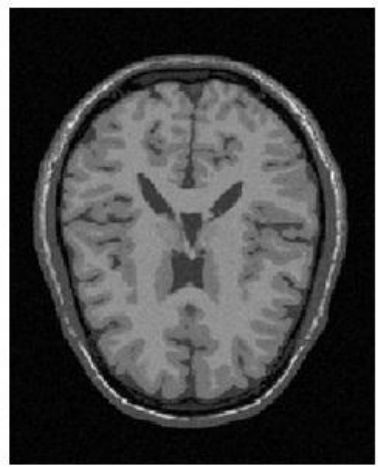

**Figure 8:** A Typical Image Available on Brain-Web Database

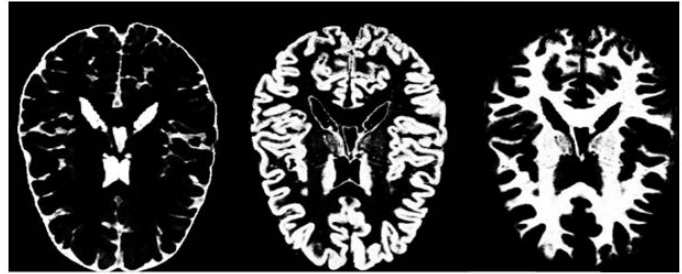

**Figure 9:** Individual Image of Gray Matter, White Matter and Cerebrospinal Fluid pertaining to Figure 8

### Exploiting Region Growth Method

To locate region grow in this study, we start from an initial point of image and the average of initial point is considered as the average of region and initial standard deviation is set to zero. We considered 8 neighbors for ini-tial point. Figure 10 shows neighboring points (red) around initial point (green). Searching is performed so that starting from initial point, neighboring pixels are checked and if they belong to that class, they are added. By adding the points to that class, the average and standard deviation ($\mu N$) are updated iteratively using the following equations.

$$\mu_N = \frac{(N-1)(\mu_N - 1) + I_N}{N} \quad (1)$$

$$V_N = \sqrt{\frac{(N-2)(V_N - 1) + I_N}{N}} \quad (2)$$

Similarly, neighbors are found for neighboring points and parameters are updated. This search continues until the first class is specified completely and no other point is added.

### Proposed Steps
Proposed steps are as following.

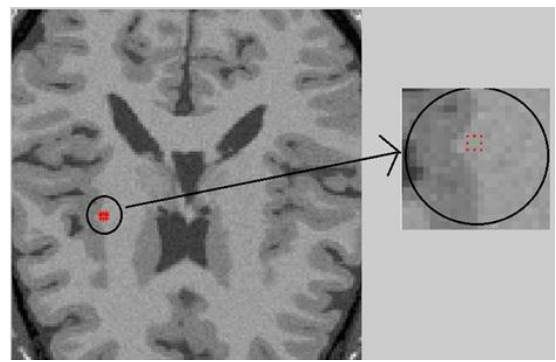

**Figure 10:** Including 8 Neighbors around Initial Point

First, initial points (white matter, cerebrospinal fluid, gray matter) specified using a simple method such as thresholding. Then, initial population for genetic algorithm is selected among these points randomly and the algorithm starts. For selected initial population, fitness function results. In this study, fitness function is defined as the difference between segmented image of database and the image that is resulted from region growth method, starting from initial random point.

$$Cost = (GM - I_1)^2 + (WM - I_2)^2 + (CSF - I_3)^2 \quad (3)$$

Where GM, WM and CSF are gray matters, white matters and cerebrospinal fluid and I1, I2 and I3 are image resulted from segmentation, respectively. Then, points are arranged with respect to fitness function. Paints processing lower values of fitness function have higher fitness resulting image to segmented image of database, more preferred. Next, more fitted chromosomes are selected for next steps. Intercourse and mutation are applied on chromosome. Next generation of chromosome are selected to continue the process and fitness is calculated for next generation. The algorithm continues to reach termination conditions (Table 2).

## Results

Aforementioned algorithm in previous section was applied to the image of Brain web database and segmented images of this database were used to calculate fitness function. Figure 11 shows the result of applying proposed algorithm on one of the images in this database. As can be seen, the proposed method has segmented the image into 3 components. Results for comparison of proposed method and original images of database are listed in Table 2. In order to compare the results of proposed algorithms, region growth method with manual selection, initial points are applied on original MRI image as well and RMS error is calculated as:

$$RMS = \frac{(\sum_{i=1}^{N} P_i - G_i)^2}{N} \quad (4)$$

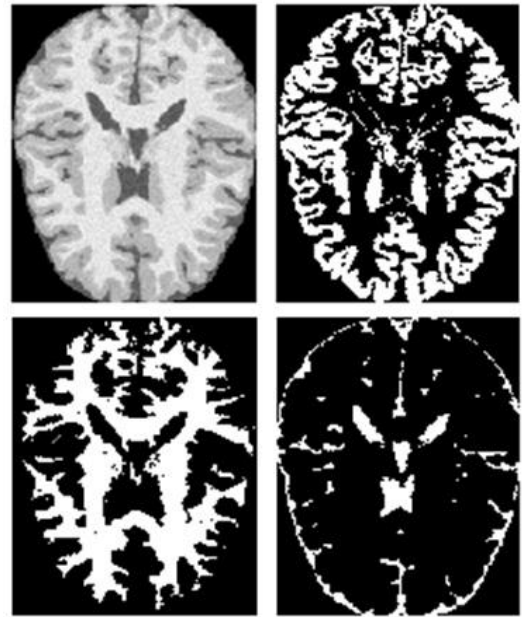

**Figure 11:** Results from Application of Proposed Algorithms on a Typical Image taken from BrainWeb Database

Where $P_i$ and $G_i$ are resulting image from proposed algorithms and segmented original image, respectively. RMS error is calculated and summed for all pixels where $P_i$ represents its pixel of image and $G_i$. Represent its pixel in the image that is well-segmented.

**Table 2:** Comparison of RMS Error in Proposed and Region Growing Method

| Comparison of Results for Image | RMS Error |
|---|---|
| Region growth method with manual selection of initial points | 0.7188 |
| Proposed method | 0.4794 |

As can be seen from these results, proposed method has reduced RMS error, meaning higher resolution.

## Discussion and Conclusion

In this study, we used region growth method to segment brain MRI images. This method includes several steps. At first, some initial points (grains) need to be selected. These points pertain to regions that need to be separated from background. Then, starting from these points and considering neighboring points, other points are checked. If points belong to the first region, based on selected similarity criterion, then they are added to that region. Initial points are selected automatically. Genetic algorithm was exploited in which after selecting initial population and defining appropriate fitness function, optimal initial points are searched. Standard deviation criterion was used to check similarity because this method has good capabilities to specify pixels belonging to a region.

Finally, proposed method was applied to brain MRI image (Figure 12). Comparing the results of proposed method and the result of region growth method with manual selection has improved brain MRI image segmentation. Improving image segmentation can greatly affect next steps for processing. It is also possible to diagnose diseases more accurately from output of image segmentation. This method is also time-saving due to automated procedure and it can be exploited for massive image segmentation, helping physicians for more accurate and fast diagnosis. The Approaches to reduce the artifacts annoying signals include the following restrictions:

The method described previously [26] based on Fourier analysis will result in removal of the original signal along with noise and it is not suitable. In the method proposed by Erfanian, et al. [33], based on simultaneous averaging, there is no good efficiency by increasing the number of signals. Kianzad et al. [34] failed to predict the estimation using artificial neural networks and this method is not effec-

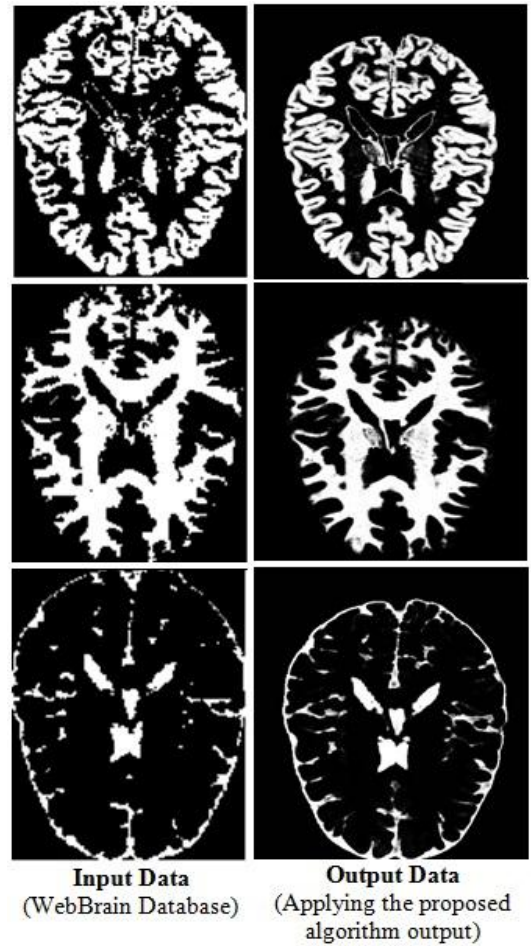

**Input Data**
(WebBrain Database)

**Output Data**
(Applying the proposed algorithm output)

**Figure 12:** Results from Application of Proposed Algorithms on a Typical Image taken from BrainWeb Database

tive by itself.

In the method suggested by Chaozhu et al. [35] and also Gholam-Hosseini et al. [36] based on adaptive filtering, some parts of signal is removed and it is not recommended according to EEG signal sequence and brain signals interpretation. Awate Method [1] does not have appropriate speed of convergence and, the algorithm process will be in trouble by additional delay. The main limitations of the method for determining the threshold is that only two classes are created in the simplest form of threshold, and it cannot be used for multi-channel images. In addition, determining the threshold would not consider the space characteristics of the image [1]. It causes method to be sensitive to non-uniformity of noise intensity that often occurs in magnetic resonance images. Both of these artifacts mainly defect the image histogram and make the segmentation operations more difficult.

The problem of Clustering Method is the time required for convergence and noise sensitivity and some researches have been done in order to address these problems [24].

In Stated method based on the area growth, the calculations are time-consuming and if the image intensity changes too much in its constituent parts, this method may lead to additional segmentation or holes in the image.

The proposed method is described to overcome the above problems. Neural networks and fuzzy inference systems are perfect nonlinear estimator that can be used. One of the efficient methods that can be applied for this purpose is the combination of neural network and fuzzy system called neural-fuzzy network. As a result, an adaptive fuzzy inference network, an adaptive network based on RBFN and a FLN-RBFN network are used. The area growth method is used for segmentation of brain images. This method is fairly simple and it is possible to choose the initial seeds and similarity criterion in it and achieve optimal segmentation. The approach used in this method is automatic and the initial seeds are selected automatically. In order to select the initial seeds of brain areas, gray matter, white matter and cerebrospinal fluid, in the area growth an automatic method is used to determine the initial seed.

Finally, it can be seen that FLN-RBFN network has better performance in removal of artifacts than ANFIS and RBFN. The method used to remove the artifacts must be such that it does not erase the original information of the signal as much as possible and solve the artifacts. According to a comparison conducted on MSE of these methods, FLN-RBFN has the least MSE i.e. the best performance. The advantage of this method is that you do not need to specify some parameters such as the threshold level in wavelet transform that heavily affect the performance of noise, and it detects the artifacts effect in an adaptive manner and removes it. The disadvantage of FLN-RBFN method is that in addition to the original signal it needs to record the artifact signal along with the original signal.

## Acknowledgments

This research has been supported by the Neuroscience Research Center, Baqiyatallah University of Medical Sciences, Tehran, Iran.

There is no actual or potential conflict of interest regarding this article.

## Conflict of Interest

There is no actual or potential conflict of interest regarding this article.

## References

1. Awate SP, Tasdizen T, Foster N, Whitaker RT. Adaptive Markov modeling for mutual-information-based, unsupervised MRI brain-tissue classification. *Med Image Anal.* 2006;**10**:726-39. doi.org/10.1016/j.media.2006.07.002. PubMed PMID: 16919993.

2. Javadpour A, Mohammadi A. Implementing a Smart Method to Eliminate Artifacts of Vital Signals. *J Biomed Phys Eng.* 2015; **5**(4): 199–206. PMCID: PMC4681465

3. Dauwels J, Vialatte F, Cichocki A. Diagnosis of Alzheimer's disease from EEG signals: where are we standing? *Curr Alzheimer Res.* 2010;**7**:487-505. doi.org/10.2174/156720510792231720. PubMed PMID: 20455865.

4. Eeckman F. Computation in neurons and neural systems: Springer Science & Business Media; 2012.

5. Saad NM, Abu-Bakar S, Muda S, Mokji M, Abdullah A, editors. Automated Region Growing for Segmentation of Brain Lesion in Diffusion-weighted MRI. Proceedings of the International MultiConference of Engineers and Computer Scientists; 2012.

6. Banerjee S. The macroeconomics of dementia-will the world economy get Alzheimer's disease? *Arch Med Res.* 2012;**43**:705-9. doi.org/10.1016/j.arcmed.2012.10.006. PubMed PMID: 23085453.

7. Xu W, Ferrari C, Wang HX. Epidemiology of Alzheimer's Disease: INTECH Open Access Publisher; 2013.

8. Li C, Xu C, Anderson AW, Gore JC, editors. MRI tissue classification and bias field estimation based on coherent local intensity clustering: A unified energy minimization framework. Information Processing in Medical Imaging: Springer; 2009. p. 288-99.

9. Xu R, Ohya J, Luo L. Segmentation of Brain MRI: INTECH Open Access Publisher; 2012.

10. Worth AJ, Makris N, Caviness Jr VS, Kennedy DN. Neuroanatomical segmentation in MRI: technological objectives. *International Journal of Pattern Recognition and Artificial Intelligence.* 1997;**11**:1161-87. doi.org/10.1142/S0218001497000548.

11. Song Z, Tustison N, Avants B, Gee JC. Integrated graph cuts for brain MRI segmentation. Medical Image Computing and Computer-Assisted Intervention–MICCAI 2006: Springer; 2006. p. 831-8.

12. Ho BC. MRI brain volume abnormalities in young, nonpsychotic relatives of schizophrenia probands are associated with subsequent prodromal symptoms. *Schizophr Res.* 2007;**96**:1-13. doi.org/10.1016/j.schres.2007.08.001. PubMed PMID: 17761401. PubMed PMCID: 2222920.

13. Slezak D, Pal S, Kang BH, Gu J, Kuroda H, Kim TH. Signal Processing, Image Processing and Pattern Recognition: International Conference, SIP 2009, Held as Part of the Future Generation Information Technology Conference, FGIT 2009, Jeju Island, Korea, December 10-12, 2009. Proceedings: Springer; 2010.

14. Hojjatoleslami S, Kruggel F, von Cramon D, editors. Segmentation of white matter lesions from volumetric MR images. Medical Image Computing and Computer-Assisted Intervention–MICCAI'99: Springer; 1999. p. 52-61.

15. Khoo VS, Dearnaley DP, Finnigan DJ, Padhani A, Tanner SF, Leach MO. Magnetic resonance imaging (MRI): considerations and applications in radiotherapy treatment planning. *Radiother Oncol.* 1997;**42**:1-15. doi.org/10.1016/S0167-8140(96)01866-X. PubMed PMID: 9132820.

16. Dunn JC. A Fuzzy Relative of the ISODATA Process and Its Use in Detecting Compact Well-Separated Clusters. *Journal of Cybernetics.* 1973;**3**:32-57. doi.org/10.1080/01969727308546046.

17. Bezdek JC. Pattern recognition with fuzzy objective function algorithms: Springer Science & Business Media; 2013.

18. Krishnapuram R, Keller JM. The possibilistic c-means algorithm: insights and recommendations. *Fuzzy Systems, IEEE Transactions on.* 1996;**4**:385-93. doi.org/10.1109/91.531779.

19. Zhang JS, Leung YW. Improved possibilistic c-means clustering algorithms. *Fuzzy Systems, IEEE Transactions on.* 2004;**12**:209-17. doi.org/10.1109/TFUZZ.2004.825079.

20. Lu L, Li M, Zhang X, editors. An improved MR image segmentation method based on fuzzy c-means clustering. Computational Problem-Solving (ICCP), 2012 International Conference on; 2012: IEEE.

21. Sikka K, Sinha N, Singh PK, Mishra AK. A fully automated algorithm under modified FCM framework for improved brain MR image segmentation. *Magn Reson Imaging.* 2009;**27**:994-1004. doi.org/10.1016/j.mri.2009.01.024. PubMed PMID: 19395212.

22. Khalid NEA, Ibrahim S, Manaf M, Ngah UK, editors. Seed-based region growing study for brain abnormalities segmentation. Information Technology (ITSim), 2010 International Symposium in; 2010: IEEE. doi.org/10.1109/itsim.2010.5561560.

23. Rose J, Grenier T, Revol-Muller C, Odet C, editors. Unifying variational approach and region growing segmentation. European Signal Processing Conference; 2010.

24. Tamilarasi M, Duraiswamy K, editors. Genetic based Fuzzy Seeded Region Growing Segmentation for diabetic retinopathy images. Computer Communication and Informatics (ICCCI), 2013 International Conference on; 2013: IEEE. doi.org/10.1109/iccci.2013.6466117.

25. Węgliński T, Fabijańska A, editors. Brain tumor segmentation from MRI data sets using region growing approach. Perspective Technologies and Methods in MEMS Design (MEMSTECH), 2011 Proceedings of VIIth International Conference on; 2011: IEEE.

26. Lehmann TM, Gonner C, Spitzer K. Survey: interpolation methods in medical image processing. *IEEE Trans Med Imaging.* 1999;**18**:1049-75. doi.org/10.1109/42.816070. PubMed PMID: 10661324.

27. Sahoo PK, Soltani S, Wong AK. A survey of thresholding techniques. *Computer vision, graphics, and image processing.* 1988;**41**:233-60. doi.org/10.1016/0734-189X(88)90022-9.

28. Rostami MT, Ghasemi J, Ghaderi R, editors. Neural network for enhancement of FCM based brain MRI segmentation. Fuzzy Systems (IFSC), 2013 13th Iranian Conference on; 2013: IEEE. doi.org/10.1109/ifsc.2013.6675661.

29. Vincent L, Soille P. Watersheds in digital spaces: an efficient algorithm based on immersion simulations. *IEEE Transactions on Pattern Analysis & Machine Intelligence.* 1991;(6):583-98. doi.org/10.1109/34.87344.

30. Trémeau A, Tominaga S, Plataniotis KN. Color in image and video processing: most recent

trends and future research directions. *Journal on Image and Video Processing*. 2008;2008:7. doi. org/10.1155/2008/581371.

31. Gómez O, González JA, Morales EF. Image segmentation using automatic seeded region growing and instance-based learning. Progress in pattern recognition, image analysis and applications: Springer; 2007. p. 192-201.

32. BrainWeb: 20 Anatomical Models of 20 Normal Brains. Available from: http//brainweb.bic.mni. mcgill.ca/brainweb/anatomic_normal_20.

33. Erfanian A, Mahmoudi B. Real-time ocular artifact suppression using recurrent neural network for electro-encephalogram based brain-computer interface. *Med Biol Eng Comput*. 2005;**43**:296-305. doi.org/10.1007/BF02345969. PubMed PMID: 15865142.

34. Kianzad R, Montazery Kordy H. Automatic Sleep Stages Detection Based on EEG Signals Using Combination of Classifiers. *Journal of Electrical and Computer Engineering Innovations*. 2013;**1**:99-105.

35. Chaozhu Z, Siyao L, Abdullah AK, editors. A New Blind Source Separation Method to Remove Artifact in EEG Signals. Instrumentation, Measurement, Computer, Communication and Control (IMCCC), 2013 Third International Conference on; 2013: IEEE. doi.org/10.1109/imccc.2013.319.

36. Gholam-Hosseini H, Nazeran H, editors. Detection and extraction of the ECG signal parameters. Engineering in Medicine and Biology Society, 1998. Proceedings of the 20th Annual International Conference of the IEEE; 1998: IEEE. doi.org/10.1109/iembs.1998.745846.